\documentclass[superscriptaddress,showpacs,aps,twocolumn,prl,floatfix]{revtex4-1}
\usepackage{amssymb}
\usepackage{amsmath}
\usepackage[dvips]{graphicx}
\usepackage{bm}

\newcommand{\bea}{\begin{eqnarray}}
\newcommand{\eea}{\end{eqnarray}}

\def\beq{\begin{equation}}
\def\eeq{\end{equation}}

\begin{document}
\title{Interplay between quantum interference and Kondo effects in non-equilibrium transport through
nanoscopic systems}
\author{P. Roura-Bas}
\affiliation{Dpto de F\'{\i}sica, Centro At\'{o}mico Constituyentes, Comisi\'{o}n
Nacional de Energ\'{\i}a At\'{o}mica, Buenos Aires, Argentina}
\author{L. Tosi}
\affiliation{Centro At\'omico Bariloche and Instituto Balseiro, Comisi\'on Nacional de
Energ\'{\i}a At\'omica, 8400 Bariloche, Argentina}
\author{A. A. Aligia}
\affiliation{Centro At\'omico Bariloche and Instituto Balseiro, Comisi\'on Nacional de
Energ\'{\i}a At\'omica, 8400 Bariloche, Argentina}
\author{K. Hallberg}
\affiliation{Centro At\'omico Bariloche and Instituto Balseiro, Comisi\'on Nacional de
Energ\'{\i}a At\'omica, 8400 Bariloche, Argentina}
\date{\today }

\begin{abstract}

We calculate the finite temperature and non-equilibrium electric current through systems described 
generically at low energy by a singlet and \emph{two}
spin doublets for $N$ and  $N \pm 1$ electrons respectively, coupled asymmetrically to two
conducting leads, which allows for destructive
interference in the conductance.
 The model is suitable for studying transport in a great variety of systems such us aromatic molecules, 
different geometries of quantum dots and rings with applied magnetic flux.
As a consequence of the interplay between interference and 
Kondo effect, 
we find changes by several orders of magnitude in the values of the conductance and its temperature
dependence as the doublet level splitting is changed by some external parameter.
The differential conductance at finite bias is negative for some parameters.

\end{abstract}

\maketitle

Transport properties of single molecules are being extensively studied
due to their potential use as active components of new electronic devices.
Experimental results show an increased conductance
at low temperatures due to the usual spin-1/2 or spin-1
Kondo effect
and quantum phase transitions were induced changing externally controlled parameters 
\cite{park,lian,roch,parks,serge}.
Measurements through single $\pi $-conjugated molecules \cite
{park,venk,dani,dado} stimulated further theoretical work.  
In particular, the possibility of controlling molecular electronics using quantum interference effects in annulene
molecules has been recently proposed \cite
{carda,ke,bege,mole}. 
The conductance depends on which sites of the molecule are
coupled to the leads and on interference phenomena related to the symmetry of the system. For certain
conditions it can be totally suppressed and restored again by
symmetry-breaking perturbations \cite{mole}.

Interference phenomena are also 
well known in rings threaded by a magnetic flux.
While sizable fluxes are at present impossible to apply to small annulene
molecules due to their small area, Aharonov-Bohm oscillations were observed
in systems involving two quantum dots (QDs).\cite{hata} Systems of three
\cite{waugh,gaud,rogge}, and more \cite{kouw} QDs have been assembled to
study the effects of interdot hopping on the Kondo effect, and other
physical properties driven by strong correlations. It has been predicted
that the transmittance integrated over a finite energy window \cite
{jagla,frie,hall} or the conductance through a ring of strongly
correlated one-dimensional systems \cite{hall,rinc} 
display dips as a function of the applied magnetic flux at \emph{fractional}
values of the flux quantum, due to spin-charge separation. While the energy
integration mimics a finite bias voltage $V$ or temperature $T$ applied to
the system, the calculations were actually performed at $V=T=0$. Moreover,
as in calculations of transport through molecules which included the effect
of correlations \cite{bege,mole,hett}, the leads were included
perturbatively, missing the Kondo regime, for which the conductance is
usually the highest \cite{soc}.
The Kondo effect was also missed in previous studies of interference effects 
at equilibrium ($V=0$) for three dots at $T>T_{K}$ \cite{kuz}, 
where $T_{K}$ is the Kondo temperature of the system, 
and in the spinless case \cite{meden}.
Experimentally, the crossing of levels and the ensuing destructive interference
has been induced applying magnetic field to a system with large, level-dependent $g$ factors \cite{nils}.

\begin{figure}[tbp]
\includegraphics[width=6.5cm]{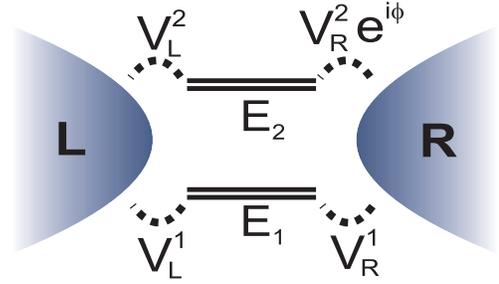}
\caption{Scheme of the relevant matrix elements at low energies.}
\label{scheme}
\end{figure}

An analysis of the current through finite rings shows
that interference phenomena take place when two levels become
degenerate \cite{rinc}. In the case of annulene molecules these states
correspond to two doublets with total wave vectors $\pm K$, degenerate due
to reflection symmetry (in absence of an external flux) \cite
{bege,mole,rinc}. For the infinite-$U$ Hubbard model, the interfering states differ in
the spin quantum numbers \cite{rinc}.

In this work we study the non-equilibrium transport in a generalized
Anderson model, whose localized part consists of a singlet with even $N$ electrons and two
doublets with $N\pm 1$ electrons (see Fig. \ref{scheme}). The model describes the low-energy
physics and interference phenomena of a wide range of physical systems, such as those mentioned above.
This calculation is a substantial improvement to previous results on conductance through interacting systems 
with interfering effects, \cite{mole,jagla,frie,hall,rinc} 
since it introduces the relevant Kondo regime neglected in those studies as well as finite 
temperature and non-equilibrium effects at finite gate voltages. 

Due to its reliability, for this study we use the non-crossing approximation (NCA) \cite{nca,nca2}. For
the case of one doublet, comparison of NCA with numerical
renormalization group (NRG) results \cite{compa}, shows that the NCA describes
accurately the Kondo physics and in addition, allows us to reach finite
bias voltages and temperatures. 
The leading behavior of the differential conductance with voltage is the correct one and $G(T)/G(0)$ 
practically coincides with the NRG result over several decades of temperature \cite{roura}.
The results should improve with increasing degeneracy (small or zero level splitting in our case) and in fact
experimental observations near quantum phase transitions were well reproduced by the NCA \cite{serge,nca2}.
We calculate the effects of interference on non-equilibrium
conductance in a regime of gate voltages for which the
doublets are favored, leading to Kondo 
effect 
and high values of the
conductance. In this approximation we asume that other higher-lying levels are negligible for the range of  parameters considered.  

The effective Hamiltonian is
\begin{eqnarray}
H &=&E_{s}|0\rangle \langle 0|+\sum_{i\sigma }E_{i}|i\sigma \rangle \langle
i\sigma |+\sum_{\nu k\sigma }\epsilon _{\nu k}c_{\nu k\sigma }^{\dagger
}c_{\nu k\sigma }  \notag \\
&&+\sum_{i\nu k\sigma }(V_{\nu }^{i}|i\sigma \rangle \langle 0|c_{\nu
k\sigma }+\mathrm{H.c}.),  \label{ham}
\end{eqnarray}%
where the singlet $|0\rangle $ and the two doublets $|i\sigma \rangle $
($i=1,2$; $\sigma =\uparrow $ or $\downarrow $) denote the localized states,
$c_{\nu k\sigma }^{\dagger }$ create conduction states in the left ($\nu =L$)
or right ($\nu =R$) lead, and $V_{\nu }^{i}$ describe the hopping elements
between the leads and both doublets, assumed independent of $k$. By a
gauge transformation, three of the four $V_{\nu }^{i}$ can be made real
and positive. If the system is a ring with the leads $\nu $ connected  at sites
$j_{\nu }$, the remaining effective hopping can be chosen as
$V_{R}^{2}=|V_{R}^{2}|e^{-i\phi }$, where $\phi =(j_{R}-j_{L})(K_{2}-K_{1})$,
and $K_{i}$ is the wave vector of $|i\sigma \rangle $ \cite{rinc}. The
magnitude of the hoppings is given by $|V_{\nu }^{i}|=|t_{\nu }\langle
i\sigma |d_{j_{\nu }\sigma }^{\dagger }|0\rangle |$, where $t_{\nu }$ is the
hopping between site $j_{\nu }$ and the lead $\nu $, and $d_{j\sigma
}^{\dagger }$ creates an electron (or a hole depending on the sign of the
gate voltage) at site $j$ \cite{ihm}. If both doublets are related by a
symmetry operation in the absence of an applied magnetic flux (for example
reflection symmetry), then $|V_{\nu }^{1}|=|V_{\nu }^{2}|$. If, in addition, $\phi =\pi $ and
$E_{2}=E_{1}$, the state $|L\sigma \rangle =(|1\sigma \rangle +|2\sigma
\rangle )/\sqrt{2}$ ($|R\sigma \rangle =(|1\sigma \rangle -|2\sigma \rangle
)/\sqrt{2}$) mixes only with the left (right) lead and therefore 
current cannot be transported between the leads.

As a basis for our study, we start from this situation of perfect
destructive interference
($\phi =\pi $) 
and also assume for simplicity symmetric coupling to
the leads, so that $|V_{\nu }^{i}|=V^{\prime }/\sqrt{2}$ and
allow for a finite splitting of the doublets $\delta =E_{2}-E_{1}$
(introduced for example by an applied magnetic flux \cite{rinc,kuz}).
For $\delta =0$, Eq. (\ref{ham}) takes the form of an SU(4) Anderson model (with
on-site hybridization $V^{\prime }$), which was used to interpret transport
experiments in carbon nanotubes \cite{cnano}. However, the system represented
is different and the conductance vanishes in our case. An important
technical difference is that in our model, the hybridization matrices of the
states $|i\sigma \rangle $ with the left and right leads are not
proportional 
for $\phi \neq 0$, and as a consequence tricks used to relate
the conductance at $V=0$ with the spectral density of states 
cannot be used and we have to calculate the conductance $G=dI/dV$ by
numerical differentiation of the current even for $V\rightarrow 0$, using a
non equilibrium formalism \cite{meir}. Changing basis $c_{1k\sigma }^{\dagger
}=(c_{Lk\sigma }^{\dagger }+c_{Rk\sigma }^{\dagger })/\sqrt{2}$, $%
c_{2k\sigma }^{\dagger }=(c_{Lk\sigma }^{\dagger }-c_{Rk\sigma }^{\dagger })/\sqrt{2}$
in Eq. (\ref{ham}), it is seen that $\delta$ acts as a symmetry
breaking field on the SU(4) Anderson model, reducing the symmetry to SU(2).
For $\delta \rightarrow \infty$, the doublet with energy $E_2$ can be neglected
and the model reduces to the usual one-level SU(2) Anderson model.
Therefore, the model interpolates between the one-level SU(4) and SU(2) Anderson models.

At $T=V=0$, the conductance $G(T,V)$ 
is given in terms of
the scattering phase
shifts in a Fermi liquid description \cite{pust}. In turn, these phase
shifts can be related to the expectation values $n_{i\sigma }=\langle
|i\sigma \rangle \langle i\sigma |\rangle $ generalizing the Friedel sum
rule \cite{lang} to the SU(4) model with a symmetry breaking field. For
constant density of states of the leads we obtain
\begin{equation}
G_{0}=G(0,0)=\frac{e^{2}}{h}\sum_{\sigma }\sin ^{2}\left[ \pi (n_{1\sigma}
-n_{2\sigma })\right] .  \label{g0}
\end{equation}

\begin{figure}[tbp]
\includegraphics[width=8.0cm]{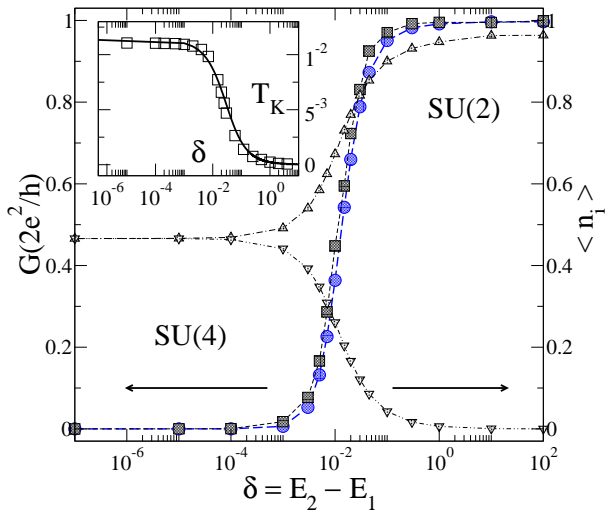}
\caption{Equilibrium conductance and $T \ll T_K$ (left scale) and occupation numbers (right scale)
as a function of the level splitting. Squares were obtained by numerical differentiation
of the current and circles correspond to Eq. (\ref{g0}). 
Inset: Kondo temperature as a function of $\delta$. Squares (solid line) correspond to the NCA (analytical) result.
Parameters are
$\Delta =0.5$. $D=10$, $E_1=-4$, $E_2=E_1+\delta$.}
\label{cond}
\end{figure}

For the numerical calculations, we assume a constant density of states per spin of
the leads $\rho $ between $-D$ and $D$ and take $\Gamma =2\Delta =2\pi \rho
(V^{\prime })^{2}$ as the unit of energy. Without loss of generality, we take
the Fermi level $\epsilon _{F}=E_{s}=0$, and assume $\delta =E_{2}-E_{1}>0$.
We also limit our present study to gate voltages
that drive the system to the Kondo regime $E_{s}-E_{1}\gg \Delta $, for which the
conductance is highest. In Fig. \ref{cond} we compare Eq. (\ref{g0}) as a
function of $\delta $, with the result obtained
by numerical differentiation of the current calculated at very low temperatures
(specifically $T=0.05T_{K}(\delta)$, see below). The
agreement is very good and gives confidence on the numerical procedure and
on the consistency of the NCA results for the current and occupation numbers 
$n_{i}=2n_{i\sigma}$, also displayed in Fig. \ref{cond}.
 We define the Kondo temperature $T_{K}$ as the half width at half maximum of
the peak nearest to the Fermi energy of the spectral density  $\rho
_{1\sigma }(\omega )$ of the lowest doublet. 
As it is apparent from the inset in Fig. \ref{cond}, we find
that to a high
degree of accuracy $T_{K}=fT_{K}^{V}$, where $f$ is a factor of the order of
1 (0.606 for the parameters of Fig. \ref{cond}), and $T_{K}^{V}$ is given by
the following expression

\begin{equation}
T_{K}^{V}=\left\{ (D+\delta )D\exp \left[ \pi E_{1}/(2\Delta )\right]
+\delta ^{2}/4\right\} ^{1/2}-\delta /2,  \label{tk}
\end{equation}%
obtained minimizing the energy of the simple variational wave function 
\begin{equation}
|\psi \rangle =\alpha |s\rangle +\sum_{ik\sigma }\beta _{ik}|i\sigma
\rangle \langle 0|c_{ik\sigma }|S\rangle ,  \nonumber
\end{equation}%
where $|S\rangle $ is the many-body singlet state with the filled Fermi sea
of conduction electrons and the state $|0\rangle $ at the localized site.

Eq. (\ref{tk}) interpolates between both one-level SU(N) limits,
$T_{K}^{V}=D\exp \left[ \pi E_{1}/(N\Delta )\right] $ ($N=4$ for $\delta =0$
and $N=2$ for $\delta \rightarrow +\infty $). 
Roughly, $T_{K}(\delta )$ remains constant at the SU(4) value  $T_{K}(0)$ ($0.016$ for the
parameters of Fig. \ref{cond}) as long as $\delta <$ $T_{K}(0)$, and then
decreases nearly exponentially (by almost two orders of magnitude for
$\delta =1$) before flattening at the one-level SU(2) value. 
As it is apparent in Fig.
\ref{cond}, $T_{K}(0)$ is also the characteristic energy scale for the
variation of the conductance with $\delta $, for $T=V=0$. For $\delta $ one
order of magnitude less than $T_{K}(0)$, $G_{0}$ is very small, while for
$\delta \gg T_{K}(0)$, $G_{0}$ approaches the ideal value
for one SU(2) doublet, $2e^{2}/h$.

\begin{figure}[tbp]
\includegraphics[width=7.5cm]{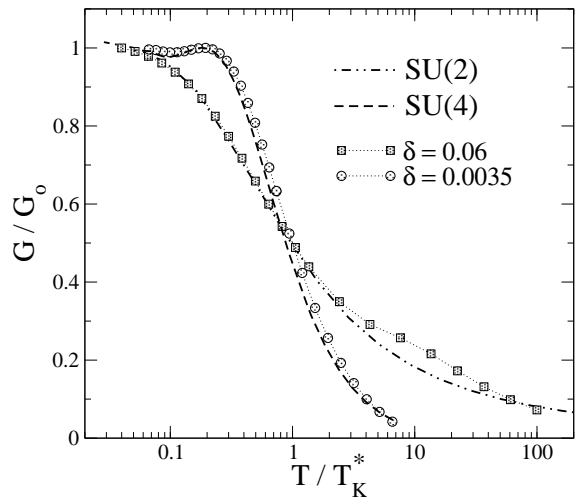}
\caption{Equilibrium conductance as a function of temperature for several level splittings.
Dashed [dashed-dot] line corresponds to the limit of one-level SU(4) [SU(2)] Anderson model.
Other parameters as in Fig. \ref{cond}. For $\delta=0.0035$, $G_{0}=0.08$ and $T_{K}^*=0.0137$,
while for $\delta=0.06$, $G_{0}=0.97$ and $T_{K}^*=0.0025$.}
\label{temp}
\end{figure}

In Fig. \ref{temp}, we show the evolution with temperature of the
conductance for several values of the level splitting  $\delta$. Note that
the ordinate axis is normalized by  $G_{0}$ and the abscissa by $T_{K}^*$,
the value of the temperature for which $G(T)=G(0)/2$. As expected $T_{K}^* \sim T_K$.
Both $G_{0}$ and $T_{K}^*$  have a strong variation with $\delta$. Near the SU(4) limit
$\delta \rightarrow 0$, $G_{0}$ is much smaller, but $T_{K}$ is much larger;
therefore $G(T)$ persists nearly constant for a much wider range of
temperatures. Instead, in the one-level SU(2) limit of large $\delta $, $G(T)$ starts
near ideal values but decays much faster with temperature. Near the SU(4)
limit, the temperature dependence of the conductance is quite similar to
that calculated for the simpler geometry of carbon nanotubes with one
particle in the quantum dot \cite{cnano}, although in the present case, the
magnitude is much smaller. The bump observed in the curve for $\delta=0.06$ is due to the contribution of the excited doublet
at energy $E_2$.

\medskip
\medskip

\begin{figure}[h!]
\includegraphics[width=7.5cm]{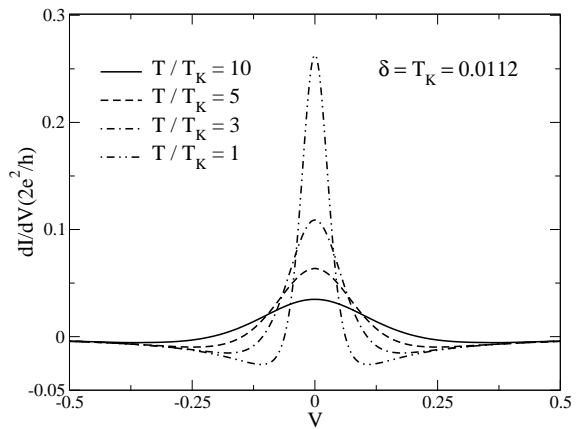}
\caption{Differential conductance as a function of bias voltage at different temperatures
for $\delta=0.0112$. Other parameters as in Fig. \ref{cond}.}
\label{didv}
\end{figure}

In Fig. \ref{didv}, we show the non-equilibrium differential conductance as a function of
bias voltage $V$ for a level splitting $\delta $ that corresponds to an
intermediate region between the SU(4) and one-level SU(2) limits ($\delta \sim
T_{K}\sim 0.01$), and for several temperatures. One observes negative $G(V)$
for voltages above the characteristic energy scale $\delta \sim T_{K}$.
Our interpretation of this result is the following. Since $\delta \sim
T_{K}$, there is only a partial effect of the interference at small
bias voltages, and the lowest lying doublet plays a dominant role. However as the
energy of the bias voltage $eV$ increases beyond the
level splitting $\delta$, both doublets contribute with nearly equal weight to the current,
but in opposite ways
due to destructive interference. Therefore, the current starts to decrease as the
effect of the excited doublet increases for larger applied bias voltages.

\begin{figure}[tbp]
\includegraphics[width=7.5cm]{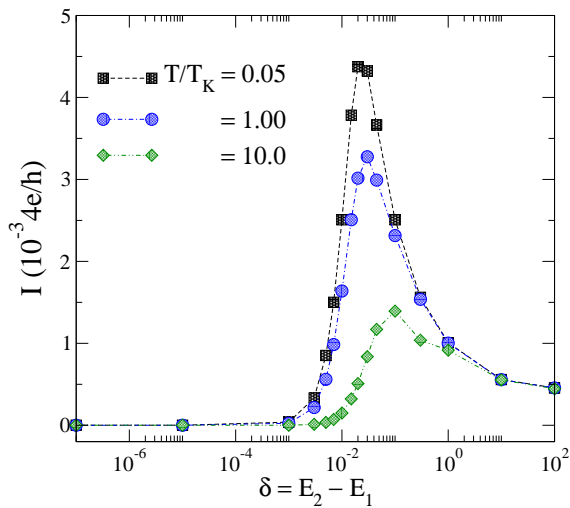}
\caption{Current as a function of level splitting $\delta$ for $eV=T_K^4=0.016$ and different 
temperatures proportional to $ T_K(\delta)$.
Other parameters as in Fig. \ref{cond}.}
\label{dip}
\end{figure}

In Fig. \ref{dip}, we display the current as a function of $\delta$ for a fixed bias voltage
at low temperatures. The bias voltage was chosen to correspond to
the characteristic energy $T_K(0)$ of the SU(4) limit.
Remarkably, a non monotonic behavior of the current is obtained due to two competing effects.
One would expect a monotonous increase of the current with the level splitting
as a consequence of the weakening of the destructive interference.
However, the Kondo temperature $T_K$ decreases strongly
with $\delta$, and for a fixed bias voltage a reduction of $T_K$ implies a decrease
in the current: at low $V$ and $T=0$, for one level with SU(N) symmetry,
$G(V)=dI/dV$ has a peak of width of the order of $2 T_K/e$. Therefore the current
at bias voltages which exceed a few $T_K/e$ is roughly proportional to $T_K$.

We have studied the finite temperature and non equilibrium transport properties of an 
effective impurity Anderson model containing two doublets, which
describes the low-energy physics of quantum interference in nanoscopic systems.
In the Kondo regime, dramatic changes in the values of the conductance and its temperature
dependence take place as the doublet level splitting is changed by some external parameter.
For total destructive interference, the model interpolates between the SU(4) Anderson
model when the splitting of the two doublets is $\delta=0$, and the usual SU(2) model
for large $\delta$. In the Kondo regime, while the characteristic temperature $T_K$
increases significantly towards the SU(4) limit, both, the equilibrium conductance (gate voltage $V=0$) 
and the total current at finite gate voltages
vanish due to destructive interference. For finite $V$ the total current peaks at finite $\delta$ 
due to the interplay between interference and Kondo effects and when $\delta \sim T_K$, 
the differential conductance becomes negative due to partial destructive interference.

In summary, by considering the interplay between two relevant effects, {\it i.e.}  quantum 
interference and Kondo screening, we have shown the important consequences this can have on 
transport properties through a great variety of nanoscopic and molecular systems.

This work was partially supported by CONICET, by PIP No 11220080101821 of CONICET, 
by PICT Nos 2006/483 and R1776 of the ANPCyT, Argentina.


\begin{thebibliography}{99}
\bibitem{park} J. Park \textit{et al.}, 
Nature (London) \textbf{417}, 722 (2002).

\bibitem{lian} W. Lian \textit{et al.}, 
Nature (London) \textbf{417}, 725 (2002).

\bibitem{roch} N. Roch \textit{et al.}, Nature \textbf{453}, 633 (2008).

\bibitem{parks} J. Parks et al., Science \textbf{328}, 1370
(2010).

\bibitem{serge} S. Florens et al., J. Phys. Condens. Matter \textbf{23}, 243202 (2011).

\bibitem{venk} B. Venkataraman \textit{et al.}, 
Nature (London) \textbf{442}, 904 (2006).

\bibitem{dani} A.V. Danilov \textit{et al.}, 
Nano Lett. \textbf{8}, 1 (2008).

\bibitem{dado} T. Dadosh \textit{et al.}, 
Nature (London) \textbf{436}, 677 (2005).

\bibitem{carda} D. Cardamone \textit{et al.}, 
Nano Lett. \textbf{6}, 2422 (2006).

\bibitem{ke} S.-H. Ke \textit{et al.}, 
Nano Lett. \textbf{8}, 3257 (2008).

\bibitem{bege} G. Begemann \textit{et al.}, 
Phys. Rev. B \textbf{77}, 201406(R) (2008).

\bibitem{mole} J. Rinc\'{o}n \textit{et al.},
Phys. Rev. Lett. \textbf{103}, 266807 (2009).

\bibitem{hata} T. Hatano \textit{et al.}, 
Phys. Rev. Lett. \textbf{106}, 076801 (2011).

\bibitem{waugh} F. R. Waugh \textit{et al.}, 
Phys. Rev. Lett. \textbf{75}, 705 (1995).

\bibitem{gaud} L. Gaudreau \textit{et al.}, 
Phys. Rev. Lett. \textbf{97}, 036807 (2006).

\bibitem{rogge} M. C. Rogge \textit{et al.},
Phys. Rev. B \textbf{77}, 193306 (2008).

\bibitem{kouw} L. P. Kouwenhoven \textit{et al.},
Phys. Rev. Lett. \textbf{65},361 (1990).

\bibitem{jagla} E.~A.~Jagla and C.~A.~Balseiro, 
Phys.~Rev.~Lett. \textbf{70}, 639 (1993).

\bibitem{frie} S.~Friederich and V.~Meden, 
Phys. Rev. B \textbf{77}, 195122 (2008).

\bibitem{hall} K.~Hallberg \textit{et al.},
Phys.~Rev.~Lett. \textbf{93}, 067203 (2004).

\bibitem{rinc} J. Rinc\'{o}n, \textit{et al.}, 
Phys. Rev. B \textbf{79}, 035112 (2009).

\bibitem{hett} M. H. Hettler \textit{et al.},
Phys. Rev. Lett. \textbf{90}, 076805 (2003).

\bibitem{soc} A. M. Lobos and A. A. Aligia, 
Phys. Rev. Lett. \textbf{100}, 016803 (2008).

\bibitem{kuz} T. Kuzmenko \textit{et al.}, Phys. Rev. Lett. \textbf{96}, 046601 (2006)

\bibitem{meden} V. Meden and F. Marquardt, Phys. Rev. Lett. \textbf{96}, 146801 (2006).

\bibitem{nils} H. A. Nilsson \textit{et al.}, Phys. Rev. Lett. \textbf{104}, 186804 (2010).

\bibitem{nca} N. S. Wingreen and Y. Meir, Phys. Rev. B \textbf{49}, 11040 (1994); 
M. H. Hettler \textit{et al.}, Phys. Rev. B \textbf{58}, 5649 (1998)

\bibitem{nca2} P. Roura-Bas and A. A. Aligia, J. Phys. Cond. Matt. \textbf{22}, 025602 (2010).

\bibitem{roura} P. Roura-Bas, Phys. Rev. B \textbf{81}, 155327 (2010) 

\bibitem{compa} T. A. Costi \textit{et al.},
Phys. Rev. B \textbf{53}, 1850 (1996).

\bibitem{ihm} A. A. Aligia \textit{et al.},
Phys. Rev. Lett. \textbf{93}, 076801 (2004).

\bibitem{cnano} J. S. Lim \textit{et al.},
Phys. Rev. B \textbf{74}, 205119 (2006); 
F. B. Anders \textit{et al.},
Phys. Rev. Lett. \textbf{100}, 086809 (2008); 
C. A. B\"{u}sser \textit{et al.}, Phys. Rev. B \textbf{83}, 125404 (2011).

\bibitem{meir} Y. Meir and N. S. Wingreen, 
Phys. Rev. Lett. \textbf{68}, 2512 (1992).

\bibitem{pust} M. Pustilnik and L. I. Glazman, 
Phys. Rev. Lett. \textbf{87}, 216601 (2001).

\bibitem{lang} D.C. Langreth, Phys. Rev. \textbf{150}, 516 (1966).



\end{thebibliography}
\end{document}